\documentclass[preprint,aps,nofootinbib,floatfix]{revtex4-1}
\pdfoutput=1

\usepackage{amsmath}  
\usepackage{amsbsy,amssymb,amsfonts}
\usepackage{graphicx}
\usepackage{color}
\usepackage{epsf}
\usepackage[utf8x]{inputenc}
\usepackage{hyperref}
\usepackage{multirow}
\usepackage{ulem}
\def\ie{{\em{i.e.}},}

\def\bravert{\egroup\,\vrule\,\bgroup}

\newcommand{\beq}{\begin{equation}}
\newcommand{\eeq}{\end{equation}}
\newcommand{\beqa}{\begin{eqnarray}}
\newcommand{\eeqa}{\end{eqnarray}}
\newcommand{\bea}{\begin{array}}
\newcommand{\eea}{\end{array}}

\newcommand{\bef}{\begin{figure}}
\newcommand{\ef}{\end{figure}}
\newcommand{\bc}{\begin{center}}
\newcommand{\ec}{\end{center}}
\newcommand{\bt}{\begin{table}}
\newcommand{\et}{\end{table}}
\newcommand{\btb}{\begin{tabular}}
\newcommand{\etb}{\end{tabular}}

\newcommand{\mea}{{\emph{et al.}}}

\newcommand{\au}{{\em a.u.}}

\def\etal{{\it et al.\ }}
\def\au{{\it a.u.}}
\def\PT{{${\cal{P,T}}$}}%
\begin{document}
\normalem
\title {\PT-Odd 
Interactions in Atomic {$^{129}$Xe} 
and Phenomenological Applications
}

\vspace*{1cm}

\author{Timo Fleig}
\email{timo.fleig@irsamc.ups-tlse.fr}
\affiliation{Laboratoire de Chimie et Physique Quantiques,
             IRSAMC, Universit{\'e} Paul Sabatier Toulouse III,
             118 Route de Narbonne, 
             F-31062 Toulouse, France }
\author{Martin Jung}
\email{martin.jung@unito.it}
\affiliation{
Dipartimento di Fisica, Universit\`a di Torino \& INFN, Sezione di Torino, I-10125 Torino, Italy
}
\vspace*{1cm}
\date{\today}

\vspace*{1cm}
\begin{abstract}
We calculate interaction constants for the contributions from \PT-odd scalar-pseudoscalar and tensor-pseudotensor operators to the electric dipole moment of ${}^{129}$Xe, for the first time in case of the former, using relativistic many-body theory including the effects of dynamical electron correlations. These interaction constants are necessary ingredients to relating the corresponding measurements to fundamental parameters in models of physics beyond the Standard Model.
We obtain $\alpha_{C_S} = \left( 0.71 \pm 0.18 \right) [10^{-23}\, e~\text{cm}]$ and $\alpha_{C_T}= \left( 0.520 \pm 0.049 \right) [10^{-20}\, \left<\Sigma\right>_{\text{Xe}}\, e~\text{cm}]$, respectively.
We apply our results to test a phenomenological relation between the two quantities, commonly used in the literature, and discuss their present and future phenomenological impact.
\end{abstract}

\maketitle
\section{Introduction}
\label{SEC:INTRO}
Electric Dipole Moments (EDMs) provide exceptionally sensitive tests for ${\cal{CP}}$-violating physics beyond the Standard Model (BSM). A significant non-zero result in any of the ongoing searches would constitute a major discovery. The quantitative interpretation of these measurements in terms of fundamental ${\cal{CP}}$-violating parameters is, however, rather complicated. The relations between these parameters and the measurements, which often involve complex systems like atoms or molecules, are established via a series of effective field theories, see Refs.~\cite{Ginges:2003qt,Pospelov:2005pr,Raidal:2008jk,fukuyama_review2012,deVries:2012ab,Engel:2013lsa,Bsaisou:2014oka,Chupp:2017rkp,Yamanaka:2017mef} for recent reviews. 
They require the evaluation of matrix elements on various scales of the problem, in particular on the atomic, nuclear, and hadronic levels, which in many cases involve large uncertainties. Apart from quantitative issues, there is also the problem that every system only probes one combination of these fundamental parameters. Constraining individual coefficients model-independently therefore requires the combination of several measurements, see for instance Refs.~\cite{Jung:2013mg,Jung:2013hka,Chupp:2014gka,Chupp:2017rkp,FleigJung_JHEP2018} for recent discussions. This is particularly complicated in the case of diamagnetic systems, for which there are many contributions. Disentangling them requires a corresponding number of competitive experimental measurements in different systems, which are not yet available. Apart from Mercury, which has provided the most stringent limit for a diamagnetic system for a long time (see Ref.~\cite{Heckel_Hg_PRL2016} for the most recent measurement), there has been recent progress for Xenon \cite{Allmendinger:2019jrk,Sachdeva:2019rkt,Sachdeva:2019blc,2018PhLA..382..588S,Liu:2020lhw}, which will allow for significant improvements in the mid-term future. 
We therefore consider in this article two contributions from semileptonic \PT-odd BSM operators in this system: the tensor-pseudotensor interaction, which, if present in a BSM model, is expected to give the dominant semileptonic contribution in diamagnetic systems \cite{Barr_eN-EDM_Atoms_1992},
and the scalar-pseudoscalar interaction, which is generally suppressed in diamagnetic systems, but provides an interesting relation to paramagnetic systems. We calculate both relevant interaction constants explicitly, which in the latter case is possible due to an approach recently developed for this coefficient in Mercury \cite{FleigJung_JHEP2018}. The explicit calculation of both coefficients in the same framework allows for testing phenomenological relations often employed in the literature.

The article is structured as follows: in section~\ref{SEC:THEORY} we review the formalism for the calculation of the two interaction constants. In section~\ref{SEC:APPL} we present the new calculation for Xenon and discuss the results. Phenomenological consequences are discussed in section~\ref{SEC:PHENO}, before concluding in section~\ref{SEC:CONCL}. Some details of the calculation are deferred to the Appendix.
\section{Theory}
\label{SEC:THEORY}
\subsection{Scalar-Pseudoscalar Nucleon-Electron (S-PS-ne) Interaction}

The theory and method of calculation for the S-PS-ne interaction {\it{via}} the magnetic hyperfine
interaction has been laid out in full detail in reference \cite{FleigJung_JHEP2018}. We summarize the 
main points in the following.

The zeroth-order wavefunctions are obtained by solving 
\begin{equation}
 \hat{H}^{(0)}\, \left| \psi^{(0)}_K \right> = \varepsilon_K^{(0)}\, \left| \psi^{(0)}_K \right>\,,
 \label{EQ:SEV_EQ}
\end{equation}
where $H^{(0)}$ is the atomic Dirac-Coulomb Hamiltonian \emph{including} the dipole energy due
to a homogeneous external electric field ${\bf{E}_{\text{ext}}}$, with the nucleus placed
at the origin:
\begin{eqnarray}
 \nonumber
 \hat{H}^{(0)} &:=& \hat{H}^{\text{Dirac-Coulomb}} + \hat{H}^{\text{Int-Dipole}} \\
      &=& \sum\limits^N_j\, \left[ c\, \boldsymbol{\alpha}_j \cdot {\bf{p}}_j + \beta_j c^2 
    - \frac{Z}{r_j}{1\!\!1}_4 \right]
        + \sum\limits^N_{j,k>j}\, \frac{1}{r_{jk}}{1\!\!1}_4
        + \sum\limits_j\, {\bf{r}}_j \cdot {\bf{E}_{\text{ext}}}\, {1\!\!1}_4\,,
 \label{EQ:HAMILTONIAN}
\end{eqnarray}
where the indices $j,k$ run over $N$ electrons, $Z$ is the proton number ($N=Z$ for neutral atoms), and
$\boldsymbol{\alpha},\beta$ are standard Dirac matrices.
We use atomic units (\au{}) throughout ($e = m_0 = \hbar = 1$).
The states $\left| \psi^{(0)}_K \right>$ are represented by electronic configuration interaction (CI) 
vectors.

The magnetic hyperfine (HF) interaction,
\begin{equation}
 \hat{H}_{\text{HF}} = - \frac{1}{2c\, m_p}\, \frac{\mu \boldsymbol{I}}{I}\,
      \cdot \sum\limits_{i=1}^n\,
          \frac{\boldsymbol{\alpha}_i \times \boldsymbol{r}_i}{r_i^3} \,,
 \label{EQ:HAM_HYPFIN}
\end{equation}
where $\mu = gI$ is the nuclear magnetic moment, $g$ the nuclear $g$-factor, $m_p$ the proton mass
and $\boldsymbol{I}$ the nuclear spin,
perturbs this wavefunction. To first order, the perturbed wavefunction can be written as
\begin{equation}
 \label{EQ:HYPFIN_PERTSUM1}
 \left| \psi^{(1)}_J \right> = \left| \psi^{(0)}_J \right>
           + \sum\limits_{K \ne J}\, \frac{ \left< \psi^{(0)}_K \right| \hat{H}_{\text{HF}}
                                           \left| \psi^{(0)}_J \right> }
           { \varepsilon_J^{(0)} - \varepsilon_K^{(0)} }\,
           \left| \psi^{(0)}_K \right>.
\end{equation}
The required matrix elements of the hyperfine Hamiltonian are obtained from
\begin{equation}
\left(A_{zk}\right)_{MN} = - \frac{\mu[\mu_N]}{2c I m_p}\, \sum\limits_{i=1}^n\, 
 \left< \psi^{(0)}_M \right| \left( \frac{\boldsymbol{\alpha}_i \times \boldsymbol{r}_i}{r_i^3} \right)_k
 \left| \psi^{(0)}_N \right>\,,
\end{equation}
where $k$ is a cartesian component and the nuclear magnetic moment enters in units of the nuclear magneton
$\mu_N = \frac{1}{2cm_p}$ (in \au).

The scalar-pseudoscalar atomic interaction constant can be defined starting from an effective theory at the nucleon level.
The corresponding Lagrangian density for the interaction of an electron with proton and neutron, $N=p,n$, can be formulated as \cite{Flambaum_Khriplovich1985}
\begin{equation}
 \label{EQ:S-PS_LAGR}
 {\cal{L}}_{\text{S-PS-ne}} = -\frac{G_F}{\sqrt{2}} \sum\limits_N\, C_S^N\,
	{\overline{\psi}}_N \psi_N\, {\overline{\psi}}\, \imath \gamma^5\, \psi
\end{equation}
where $G_F$ is the Fermi constant and $\psi_{(N)}$ are field operators. The Wilson coefficients $C_S^N$ are determined by ${\cal{CP}}$-odd interactions at higher energies. 
%
%
%
%
Summing over electrons and nucleons, this yields for the effective interaction in a many-electron atom
\begin{equation}
 \hat{H}^{\text{eff}}_{\text{S-PS-ne}} = \imath \frac{G_F}{\sqrt{2}}\, AC_S\, 
            \sum\limits_e\,\gamma^0_e\, \gamma^5_e\, \rho({\bf{r}}_e)\,,
 \label{EQ:HAM_S-PS-ne}
\end{equation}
%

where 
$A$ is the nucleon number, $\rho_N$ the normalized nuclear charge density, and $\gamma^{\mu}$ are standard Dirac matrices. $C_S$ is the dimensionless S-PS-ne coupling constant, given as 
\begin{equation}
    C_S = \frac{Z}{A}C_S^p + \frac{A-Z}{A}C_S^n\,.
\end{equation}
Note that this combination is in principle isotope-specific, but approximately universal for heavy atoms \cite{Jung:2013mg}.
To leading order the S-PS-ne energy shift is thus written as
\begin{equation}
    \left(\Delta \varepsilon\right)_J = 
        \frac{1}{\langle \psi^{(1)}_J|\psi^{(1)}_J\rangle}\left<
	    \hat{H}^{\text{eff}}_{\text{S-PS-ne}}\right>_{\psi^{(1)}_J}.
	    \end{equation}
Finally, the atomic EDM due to S-PS-ne interaction is
\begin{equation}
 d_a = \alpha_{C_S}C_S
\end{equation}
where the atomic interaction constant is determined from
\begin{equation}
 \alpha_{C_S}(\psi_J) = 
 \frac{-A \frac{G_F}{\sqrt{2}}}{E_{\text{ext}}\, \left< \psi^{(1)}_J \right| \left. \psi^{(1)}_J \right>}\,
 \left[ \sum\limits_{K \ne J}\, \frac{ \left< \psi^{(0)}_K \right| \hat{H}_{\text{HF}}
 \left| \psi^{(0)}_J \right> \left< \psi^{(0)}_J \right| \imath\, \sum\limits_e\,\gamma^0_e\, \gamma^5_e\, \rho({\bf{r}}_e)
        \left| \psi^{(0)}_K \right> } { \varepsilon_J^{(0)} - \varepsilon_K^{(0)} }\, + h.c. \right]
 \label{EQ:ALPHA_CS}
\end{equation}
For convenience, we use in the following also the S-PS-ne ratio $S$ (not to be confused with the nuclear
Schiff moment, also denoted $S$ in the literature), defined as
\begin{align}
 S := \frac{d_a}{A C_S \frac{G_F}{\sqrt{2}}}=\frac{\alpha_{C_S}}{A\frac{G_F}{\sqrt{2}}}
 \approx - \frac{\left< \imath \sum\limits_e\, \gamma^0_e\, \gamma^5_e\, \rho({\bf{r}}_e) 
 \right>_{\psi^{(1)}(E_{\text{ext}})}}{E_{\text{ext}}\,\langle \psi^{(1)}|\psi^{(1)}\rangle}.
\label{EQ:SRATIO}
\end{align}

\subsection{Tensor-Pseudotensor Nucleon-Electron (T-PT-ne) Interaction}

Similarly to the scalar-pseudoscalar interaction, the T-PT-ne atomic interaction constant can be defined starting from an effective 
theory at the nucleon level.
The corresponding Lagrangian density for the interaction of an electron with proton and neutron, $N=p,n$, can be formulated as
\begin{equation}
 \label{EQ:T-PT_LAGR}
 {\cal{L}}_{\text{T-PT-ne}} = \frac{1}{2} \frac{G_F}{\sqrt{2}} \sum\limits_N\, \varepsilon^{\mu\nu\rho\sigma}C_T^N\,
	{\overline{\psi}}_N {\sigma_N}_{\hspace*{-0.05cm}\mu\nu} \psi_N\, {\overline{\psi}} {\sigma}_{\hspace*{-0.05cm}\rho\sigma} \psi
\end{equation}
where $G_F$ is the Fermi constant,  ${\sigma}^{\rho\sigma} = 
\frac{\imath}{2}\left( \gamma^{\rho}\gamma^{\sigma} - \gamma^{\sigma}\gamma^{\rho} \right)$ ($\sigma_N^{\mu\nu}$ analogously), 
and $\psi_{(N)}$ are field operators.
$C_T^N$ are Wilson coefficients determined by ${\cal{CP}}$-odd interactions at higher energies.
We use the convention $\epsilon^{0123}=1$.
Based on this expression and using the identity
$\frac{1}{2} \varepsilon^{\mu\nu\kappa\lambda} {\sigma}_{\kappa\lambda} = -\imath \gamma^5 {\sigma}^{\mu\nu}$ with electronic $\gamma^5 = \imath \gamma^0 \gamma^1 \gamma^2 \gamma^3$, an effective first-quantized Hamiltonian can be written as
\begin{equation}
 \label{EQ:T-PT_HAM1}
 \hat{H}^{\text{eff}}_{\text{T-PT-ne}} = \frac{\imath G_F}{\sqrt{2}}\,\sum_N C_T^N\, \rho_N({\bf{r}})
                \gamma^0 {\sigma_N}_{\mu\nu} \gamma^5 {\sigma}^{\mu\nu}\,,
\end{equation}
where $\rho_N({\bf{r}})$ denotes the probability density of the corresponding nucleon at the position of the electron
and the electronic
Dirac matrix $\gamma^0$ originates in the field creator ${\overline{\psi}}$.
The term ${\sigma_N}_{\mu\nu} \gamma^5 {\sigma}^{\mu\nu}$ is the signature of a rank-2 Dirac nuclear-tensor electronic-pseudotensor operator, and satisfies the following identity:
\begin{equation}
 \label{EQ:T-PT_EXPRESSION}
 {\sigma_N}_{\mu\nu} \gamma^5 {\sigma}^{\mu\nu} = 2 \gamma_N^0 {\boldsymbol{\gamma}}_N \cdot {\boldsymbol{\Sigma}}
                                                + 2 \gamma^0 {\boldsymbol{\Sigma}}_N \cdot {\boldsymbol{\gamma}}\,,
\end{equation}
where ${\boldsymbol{\Sigma}}_N$ denotes the nuclear and ${\boldsymbol{\Sigma}}$ the electronic spin matrix. 
In Appendix \ref{ASEC:SIGMA} we demonstrate that the electronic expectation value $\left<\psi | {\boldsymbol{\Sigma}} | \psi\right>$ of the first term on the right-hand side of this equation is strictly zero if $\psi$ is a closed-shell wavefunction for a many-electron state with valence configuration $ns_{1/2}^2$, even including an external electric field.
This condition is lifted only by internal magnetic couplings such as the hyperfine or electronic spin-orbit interactions. The resulting contributions appear therefore only at higher orders in the perturbative series and can be neglected in the following.
The T-PT-ne Hamiltonian used here consequently stems from
the second term on the right-hand side of Eq.~(\ref{EQ:T-PT_EXPRESSION}):
\begin{equation}
 \label{EQ:T-PT_HAM_FIN}
 \hat{H}^{\text{eff}}_{\text{T-PT-ne}} = \frac{\imath G_F}{\sqrt{2}}\, \sum_N 2 C^T_N\, {\boldsymbol{\Sigma}}_N \cdot 
  {\boldsymbol{\gamma}} \rho_N({\bf{r}})\,.
\end{equation}
Up to this point the formulation is in terms of individual nucleons. We now consider this interaction in the context of a nucleus with many protons and neutrons.
Then, using Eq.~(\ref{EQ:T-PT_HAM_FIN}), we can write
\begin{equation}
 \label{EQ:T-PT_HAM_pn}
 \hat{H}^{\text{eff}}_{\text{T-PT-ne}} = \frac{2 \imath G_F}{\sqrt{2}}\, \rho(\mathbf r)\left[ C^T_p\sum_i   {\boldsymbol{\Sigma}}_i \cdot 
  {\boldsymbol{\gamma}} +  C^T_n\sum_j  {\boldsymbol{\Sigma}}_j \cdot 
  {\boldsymbol{\gamma}}\, \right]\,,
\end{equation}
where $\rho(\bf r)$ now refers to the (normalized) nuclear density which we assume to be equal for protons and neutrons, and the indices $i,j$ run over all protons and neutrons, respectively. In a nuclear shell model, the spin sums over closed shells of either protons or neutrons vanish, so the total sums will be dominated by those nucleons that do not form closed shells. Their precise values depend on the adopted nuclear model; they determine the linear combination of $C_T^{p,n}$ that is probed by a given nuclear isotope. Observing
\begin{equation}
    \langle \mathbf \Sigma\rangle_{A} = \langle \mathbf \Sigma_p^A\rangle + \langle \mathbf \Sigma_n^A\rangle\quad \mbox{and}\quad \langle\mathbf \Sigma\rangle = \langle\Sigma\rangle\, \mathbf I/I\,,
\end{equation}
we can write for the effective, isotope-specific interaction constant $C_T^A$
\begin{equation}
    \langle \Sigma \rangle_A C_T^A = \langle \Sigma \rangle_p^A C_T^p + \langle \Sigma \rangle_n^A C_T^n\,.
\end{equation}
Therefore, the Hamiltonian for electron-nucleus interaction becomes
\begin{equation}
 \hat{H}^{\text{eff}}_{\text{T-PT-ne}} = \frac{2\imath G_F}{\sqrt{2}}\,  \rho(\mathbf r)\,C_T^A\,{\langle\Sigma\rangle}_A \frac{\mathbf I\cdot \boldsymbol{\gamma}}{I}\,.
\end{equation}
In a setup with rotational symmetry around the $z$ axis, the $1,2$ components of $\langle\mathbf \Sigma\rangle$ will vanish. Considering furthermore a nuclear state $|I,M_I=I\rangle$ and integrating over the nuclear coordinates, the Hamiltonian can be written as
\begin{equation}
 \hat{H}^{\text{eff}}_{\text{T-PT-ne}} = \frac{2\imath G_F}{\sqrt{2}}C_T^A\,{\langle\Sigma\rangle}_A \gamma_3 \,  \rho(\mathbf r)\,.
\end{equation}
The evaluation of this Hamiltonian in the multi-electron environment of the given atom determines the T-PT-ne energy expectation value $R_T$: defining the matrix element
\begin{equation}
 \label{EQ:T-PT-ME}
M_{\rm e}^{\rm T-PT} = \left< \psi^{(0)}_I
                 \left| \imath \sum\limits_{j=1}^n\, \left(\gamma_3\right)_j\, \rho({\bf{r}}_j) 
                                     \right| \psi^{(0)}_I \right>\,,
\end{equation}
$R_T$ is given as
\begin{equation}
 \label{EQ:RT_DETAIL}
  R_T = \sqrt{2} G_F \left<\Sigma\right>_{A}\, M_{\rm e}^{\rm T-PT}.
\end{equation}
The expectation value $\left<\psi | \gamma_3 | \psi\right>$ for a closed-shell electronic state is shown to be non-zero in Appendix
\ref{ASEC:GAMMA}.

The atomic electric dipole moment due to a tensor-pseudotensor interaction can be written as
\begin{equation}
 d_a = C_T^A\, \alpha_{C_T}
\end{equation}
and we define the T-PT interaction constant in the quasi-linear regime (see also Ref. \cite{Fleig_Skripnikov2020}) with very small external electric fields as
\begin{equation}
    \alpha_{C_T} := \frac{R_T}{E_{\text{ext}}}.
\end{equation}

\subsection{Electric Dipole Polarizability}

The ${\cal{P,T}}$-odd \emph{permanent} EDM is measured via the atom's response to an electrical field. Hence, the reaction of an atom to such a field is a quantity of interest. Specifically, the \emph{dynamic} dipole moment of the atom, \emph{i.e.}, the induced dipole moment that is itself proportional to the electric field (and hence does not violate any fundamental symmetries), is characterized to leading order by the \emph{atomic electric polarizability} $\alpha$. This is a relatively easily measurable quantity; the comparison of its experimental value with the value calculated in a given electronic-structure model provides a measure of the quality of that model used also in the calculations of \PT-odd effects. 

A Taylor expansion of the field-dependent total electronic energy $\varepsilon$ around electric field $E=E_z=0$ reads
\begin{align}
 \varepsilon(E) &= \varepsilon^{(0)} + \left. \frac{\partial \varepsilon(E)}{\partial E} \right|_{E=0}\, E
              + \frac{1}{2}\, \left. \frac{\partial^2 \varepsilon(E)}{\partial E^2} \right|_{E=0}\, E^2 + \ldots\\
              &\equiv \epsilon^{(0)} -d_A E -\frac{1}{2}a_{zz}E^2+\ldots\,,
 \label{EQ:FDENERGY_TAYLOREXP}
\end{align}
with the atomic EDM $d_A$ and the 2nd order polarizability tensor $(\alpha_{ij})$. 

We obtain $\alpha \equiv \alpha_{zz}$ by calculating $\varepsilon(E)$ at a finite set of field points and
then determining the second derivative in Eq. (\ref{EQ:FDENERGY_TAYLOREXP}) from a fit to the perturbed 
energies.
\section{{$^{129}$Xe} EDM calculations}
\label{SEC:APPL}
\subsection{Technical details}

Atomic basis sets of Gaussian functions are used, denoted valence double-zeta (vDZ), valence
triple-zeta (vTZ), and valence quadruple-zeta (vQZ) including all available polarizing and 
valence-correlating functions \cite{4p-basis-dyall-2,dyall_tl_tz}. 
The complete sets amount to (21s 15p 11d 2f), (29s 22p 18d 7f 2g), and (34s 28p 19d 12f 7g 2h) 
functions for vDZ, vTZ, and vQZ, respectively. Wavefunctions for the {$^1S_0$} electronic ground 
state of Xe are obtained through a closed-shell Hartree-Fock (HF) calculation using the 
Dirac-Coulomb Hamiltonian including the external electric field, see Eq. (\ref{EQ:HAMILTONIAN}).
The DCHF calculation is followed by a linear expansion in the basis of Slater determinants formed by the 
occupied and virtual sets of 4-spinors and diagonalization of the DC Hamiltonian including the external 
electric field in that basis (Configuration Interaction (CI) approach) \cite{knecht_luciparII}. The 
resulting ``correlated'' wavefunctions $\psi^{(0)}_I$ -- where the CI expansion coefficients are fully 
relaxed with respect to the external electric field -- are then introduced into Eqs. 
(\ref{EQ:HYPFIN_PERTSUM1}) and (\ref{EQ:T-PT-ME}). Coupled-cluster (CC) calculations are performed
by exponentially expanding the wavefunction into the $n$-electron sector of Fock space using the same 
set of DCHF spinors as for the CI calculations.
The nomenclature for both CI and CC models is that S, D, T, etc. denote Singles, Doubles, 
Triples etc.
replacements with respect to the closed-shell DCHF determinant. The following number is the number of
correlated electrons and encodes which occupied shells are included in the CI or CC expansions. 
For the calculation of $\alpha$ and $\alpha_{C_T}$ we have 
$8 \mathrel{\widehat{=}} (5s,5p)$, 
$16 \mathrel{\widehat{=}} (4s,4p,5s,5p)$, 
$18 \mathrel{\widehat{=}} (4d,5s,5p)$, 
$24 \mathrel{\widehat{=}} (3s,3p,4s,4p,5s,5p)$, 
$26 \mathrel{\widehat{=}} (4s,4p,4d,5s,5p)$, 
$32 \mathrel{\widehat{=}} (2s,2p,3s,3p,4s,4p,5s,5p)$.
$36 \mathrel{\widehat{=}} (3d,4s,4p,4d,5s,5p)$.
The notation type S16\_SD32, as an example, means that the model SD32 has been approximated by
omitting Double excitations from the $(2s,2p,3s,3p)$ shells.
The nuclear spin quantum number is $I=1/2$ for {$^{129}$Xe} \cite{stone_INDC2015}. 
The atomic nucleus is described by a Gaussian charge distribution
\cite{Visscher_Dyall_nuclcha} with exponent $\zeta_{\text{Xe}} = 1.8030529331 \times 10^8$.
Atomic static electric dipole polarizabilites are obtained from fitting the total electronic energies using
seven points of field strengths $E_{\rm{ext}} \in \{-1.2, -0.6, -0.3, 0.0, 0.3, 0.6, 1.2\} \times 10^{-4}$
\au\ For the calculation of the T-PT interaction constant we use $E_{\rm{ext}} = 0.3 \times 10^{-4}$ \au\,,
for the S-PS-ne interaction constant $E_{\rm{ext}} = 0.5 \times 10^{-3}$ \au\

A locally modified version of the \verb+DIRAC+ program package \cite{DIRAC16} has been used for all
electronic-structure calculations.
Interelectron correlation effects are taken into account through Configuration Interaction (CI) theory
by the \verb+KRCI+ module \cite{knecht_luciparII} and Coupled Cluster (CC) theory by the \verb+RELCCSD+ 
module \cite{relccsd:parallel:1,relccsd:parallel:2} both as implemented in \verb+DIRAC+ \cite{DIRAC_pub}.

\subsection{Results and discussion}

Results for electric dipole polarizability and the atomic T-PT interaction constant are compiled in Table \ref{TAB:RT_XE}. We discuss the quantities separately.

\subsubsection{Electric dipole polarizability}

The most striking observation from CC calculations is the apparent correspondence of the CCSD(T)8 model
using the largest atomic basis set with the experimental result and the ensuing decrease of the
polarizability when outer- and inner-core shell electrons are included in the CC expansion. This same
trend is observed for the CI models as well, although here it tends to overshoot somewhat. However,
the difference between $\alpha$(CISD18) and $\alpha$(CCSD(T)18) is only about $3.5$\% in the vQZ basis. Since
the basis-set effect on the polarizability is of the order of $1-2$\%, a similar conclusion can be made
when comparing $\alpha$(CISD26) and $\alpha$(CCSD(T)26).
The latter model yields a polarizability that differs from the experimental result by about $4$\%.
Given that accounting for both perturbative triple excitations and inner-core correlations tends to
decrease $\alpha$ and the basis-set effect is so small, it is unclear how to explain this deviation.
The CC models used by Y. Singh \mea\ and Sakurai \mea\ (quoted in Table \ref{TAB:RT_XE}) yield
polarizabilities that appear to agree better with experiment. However, it is not evident that
these latter models give the right answer for the right reason, since the
relativistic CC calculation of Nakajima \mea\ that includes full iterative triple excitations yields
a smaller result, very close to the present best value from the model vQZ/CCSD(T)26.

Nonetheless, the final deviations from experiment of both CI and CC models employed in the present
work are sufficiently small for the present purpose of determining atomic ${\cal{P,T}}$-odd
interaction constants. A deeper investigation into the above issue will therefore not be attempted
here. A further reason for this will be given in the next subsection.

\subsubsection{T-PT interaction}

\begin{table}[h]
 \caption{Static electric dipole polarizability and T-PT-ne interaction constant for the Xe atom 
          using different wavefunction models
          \label{TAB:RT_XE} }

 \vspace*{0.5cm}
 \begin{tabular}{l|cc|ccc}
 Model/virtual cutoff (vDZ,vTZ,vQZ) [\au] & 
 \multicolumn{2}{c|}{$\alpha$ [\au]} &
 \multicolumn{3}{c}{$\alpha_{C_T}$ [$10^{-20}$ $\left<\Sigma\right>_{\text{Xe}}$ $e$ cm]} \\ \hline\hline
                             &  \multicolumn{2}{c|}{Basis set}  & \multicolumn{3}{c}{Basis set}     \\
                             & vTZ        & vQZ        & vDZ            & vTZ          & vQZ        \\ \hline
        RPA/-                & $27.08$    & $26.76$    & $0.460$        & $0.552$      & $0.566$    \\ 
        SD8/80,100,60        & $27.42$    & $28.58$    & $0.438$        & $0.517$      & $0.534$    \\
        SDT8/80,100,60       & $27.32$    & $28.77$    & $0.438$        & $0.514$      & $0.531$    \\
        SDTQ8/80,12,60       & $     $    & $     $    & $0.435$        & $0.510$      & $     $    \\
        CCSD8/80,100,60      & $     $    & $27.59$    & $     $        & $     $      & $     $    \\
        CCSD(T)8/80,100,60   & $     $    & $27.81$    & $     $        & $     $      & $     $    \\
        SD16/80,100,60       & $27.07$    & $26.80$    & $0.439$        & $0.518$      & $0.534$    \\
        SD8\_SDT16/80,100,60 & $28.58$    & $     $    & $0.438$        & $0.514$      & $     $    \\
        SD18/80,100,60       & $26.36$    & $26.29$    & $     $        & $0.532$      & $     $    \\
        CCSD18/80,100,60     & $27.81$    & $27.26$    & $     $        & $     $      & $     $    \\
        CCSD(T)18/80,100,60  & $     $    & $27.19$    & $     $        & $     $      & $     $    \\
	SD24/80,100,60       & $26.44$    & $26.54$    & $0.441$        & $0.520$      & $0.538$    \\
	SD26/80,100,60       & $26.32$    & $     $    & $     $        & $0.530$      & $     $    \\
        CCSD26/80,100,60     & $     $    & $27.05$    & $     $        & $     $      & $     $    \\
        CCSD(T)26/80,100,60  & $     $    & $26.86$    & $     $        & $     $      & $     $    \\
   S16\_SD32/80,100,60       & $     $    & $     $    & $     $        & $0.520$      & $     $    \\
	SD32/80,100,60       & $26.33$    & $     $    & $0.442$        & $0.521$      & $0.538$    \\
	SD36/80,100,60       & $25.58$    & $     $    & $0.453$        & $0.536$      & $     $    \\
 {\bf{vQZ/SD32/60}} +$ \boldsymbol{\Delta}$ & \multicolumn{2}{c|}{$\boldsymbol{ }$} 
   &\multicolumn{3}{c}{$\boldsymbol{0.549}$}  \\ \hline Flambaum  \mea\footnote{Ref. \cite{Flambaum_Khriplovich1985} }  
   &  \multicolumn{2}{c|}{      }  &  \multicolumn{3}{c}{$0.41$}     \\
   Dzuba \mea\footnote{Ref. \cite{dzuba_flambaum_PRA2009}} RPA 
   &  \multicolumn{2}{c|}{      }  & \multicolumn{3}{c}{$0.57$}       \\
   M{\aa}rtensson-Pendrill\footnote{Ref. \cite{mar-pendrill-_PRL1985}} RPA 
   &  \multicolumn{2}{c|}{      }  & \multicolumn{3}{c}{$0.52$}       \\           
   Nakajima \mea\footnote{Ref. \cite{nakajima_pol_2001}} RCCSDT
   &   \multicolumn{2}{c|}{$27.06$}        &                &         \\
   Y. Singh \mea\footnote{Ref. \cite{singh_sahoo_das_Xe_PRA2014}} CCSD$_p$T
   &   \multicolumn{2}{c|}{$27.78$}        &    \multicolumn{3}{c}{{$0.50$}}\\
   Sakurai \mea\footnote{Ref. \cite{sahoo_XePol_2018},\cite{Sakurai:2019vjs}} RNCCSD
   &   \multicolumn{2}{c|}{$27.51$}        &    \multicolumn{3}{c}{{$0.49$}}\\
   {Sakurai} \mea\footnote{Ref. \cite{Sakurai:2019vjs}} RCCSD(SC)
   &   \multicolumn{2}{c|}{$28.12$}        &    \multicolumn{3}{c}{{$0.48$}}\\
   Experiment\footnote{Ref. \cite{Hohm_XePol_1990}} 
   &   \multicolumn{2}{c|}{$27.815(27)$}   &                &         \\
 \end{tabular}
\end{table}

The general strategy for obtaining an accurate final result for $\alpha_{C_T}$ is to identify the leading physical
effects on $\alpha_{C_T}$ using the small basis set (vDZ). In a second step these dominant effects are determined
more accurately using the larger basis sets, and finally a correction to the final value is applied by
adding differences from effects neglected with the larger basis sets.

All of the present results contain the core contribution to $\alpha_{C_T}$ from frozen shells.
The predominant effect on $\alpha_{C_T}$ comes from electron correlations (Double (D) excitations) among the 
valence $(5s,5p)$ electrons which diminish $\alpha_{C_T}$, and the core 
electrons, which increase $\alpha_{C_T}$. Interestingly, these two contributions cancel each other to some degree. 
When comparing the correlation trends for $\alpha$ and $\alpha_{C_T}$, for the most part increasing polarizability
is correlated with decreasing T-PT interaction, and vice versa. Physically speaking, the ``softer''
the atom (\ie\, the more polarizable) the smaller its EDM due the electron-nucleon time-reversal-violating
interaction. However, this correlation is not always satisfied, for example when comparing the models
SD24 and SD26, or SD32 and SD36. On the other hand, the trend only seems to be broken when electron
shells of certain angular momenta are replaced by shells of different angular momentum, not when
shells of the same angular momentum are added to the correlation expansion.

The total effect of adding excitations from shells $n=2$ through $n=5$ is about $-5$\% on $\alpha_{C_T}$. The 
effect of electron correlations from excitations out of the $3d$ and $4d$ shells 
will be added as a correction to the final value.

Since inner-shell holes are relevant in assessing $\alpha_{C_T}$ and valence basis sets are used
in the present study, the question arises whether these basis sets are adequate for the present purpose.
Table \ref{TAB:ECORR} shows that indeed the correlation energy per correlated electron diminishes with
%
\begin{table}[h]
 \caption{Correlation energy per correlated electron with vTZ basis
          \label{TAB:ECORR} }

 \vspace*{0.5cm}
 \begin{tabular}{l|cc}
 Model/virtual cutoff [\au]  &  $\frac{E_{\text{corr}}}{N}$ $\left[ m E_H \right]$  
                             &  $\alpha_{C_T}$ [$10^{-20}$ $\left<\Sigma\right>_{\text{Xe}}$ $e$ cm]  \\ \hline
 SD8/100        & $20.0$  &  $0.517$  \\ 
 SD8/550        & $20.0$  &  $0.517$  \\
 SD16/100       & $16.1$  &  $0.518$  \\
 SD24/100       & $13.7$  &  $0.520$  \\
 SD24/550       & $15.5$  &  $0.521$  \\
 SD32/100       & $11.2$  &  $0.521$  \\
 SD32/550       & $17.0$  &  $0.525$
 \end{tabular}
\end{table}
%
the number of correlated electrons. This is to a significant part due to the truncation of
the virtual spinor space or, in other words, the lack of core-correlating functions in the employed
part of the basis set, which becomes evident when increasing the cutoff to $550$ \au, increasing the
correlation energy in a range of $10$-$50$\%. However, the property $\alpha_{C_T}$ varies by less than one
percent with increasing cutoff, thus justifying the truncation of the virtual spinor space (and
the use of a valence basis set).

From the results using the vDZ basis set valence triple excitations seem to be unimportant and valence
quadruple excitations seem to affect $\alpha_{C_T}$ by less than $1$\%. Since small basis sets do not catch the 
full effects of dynamic correlations we test these observations. 
Indeed, larger basis sets augment the effects from higher excitations, but the change is not dramatic.
Full triple and quadruple excitations from the ($5s,5p$) shells lead to a decrease of $\alpha_{C_T}$ by only
$-1.6$\%. The use of a smaller virtual space for the SDTQ8 model has been justified by comparing 
SDT8 models at $12$ \au{} and $100$ \au{} which produce the same shift on $\alpha_{C_T}$. So the effect of
higher excitations is already correctly described at the lower cutoff. Finally, combined higher 
excitations from inner shells and valence shells do not affect $\alpha_{C_T}$ notably, as the model SD8\_SDT16
demonstrates.

The final value for the tensor-pseudotensor interaction constant is obtained from the base value
calculated with the model vQZ/SD32/60 to which a shift is applied, determined as follows:
\begin{eqnarray*}
 \Delta \alpha_{C_T} && = \alpha_{C_T}({\rm{vTZ/SD36/100}}) - \alpha_{C_T}({\rm{vTZ/SD16/100}}) \\
            &+&  \alpha_{C_T}({\rm{vTZ/SDTQ8/100}}) - \alpha_{C_T}({\rm{vTZ/SD8/100}})\,.
\end{eqnarray*}
%

The uncertainty of the final value for the atomic T-PT-ne interaction constant is estimated by linearly 
adding uncertainties for relevant individual degrees of freedom in the calculations. These are $3$\% for the
atomic basis set, $4$\% for remaining inner-shell correlations and $2$\% for the higher excitation ranks
not considered by the present models, adding up to a total estimated uncertainty of about $9$\%.
We thus write the T-PT interaction constant including the estimated uncertainty as
\begin{equation}\label{eq::CTres}
 \alpha_{C_T} = \left( 0.549 \pm 0.049 \right) [10^{-20}\, \left<\Sigma\right>_{\text{Xe}}\, e\, \text{cm}].
 \end{equation}
Recent work \cite{Hubert2020} has made possible the use of the more accurate Fermi distribution for the nuclear density $\rho({\bf{r}})$ in our atomic calculations. Replacing the Gaussian distribution by this Fermi distribution in the \PT-violating operator evaluated in Eq.~(\ref{EQ:T-PT-ME}) results in a drop of $\alpha_{C_T}$ by about $5.5$\% (vQZ/SD32/60). We include this as a correction and give the final value for the T-PT interaction constant as
\begin{equation}\label{eq::CTresFermi}
 \alpha_{C_T} = \left( 0.520 \pm 0.049 \right) [10^{-20}\, \left<\Sigma\right>_{\text{Xe}}\, e\, \text{cm}].
\end{equation}
The S-PS operator in Eq. (\ref{EQ:HAM_S-PS-ne}) also depends explicitly on the nuclear density. A change to using a Fermi distribution in this case is ongoing work. However, the ratio $\frac{\alpha_{C_T}}{\alpha_{C_S}}$ will be less affected by changing the nuclear density due to cancellations. The discussion in section \ref{SEC:PHENO} will thus be based on the value from Eq.~(\ref{eq::CTres}).

\subsubsection{S-PS interaction}

S-PS nucleon-electron interaction constants -- calculated according to Eqs. (\ref{EQ:ALPHA_CS}) and
(\ref{EQ:SRATIO}) -- are shown in Table \ref{TAB:129Xe:AEHYEN}. The general pattern in all of the 
different models is the appearance of two major contributions in the sum over states, see Eq.~(\ref{EQ:ALPHA_CS}). These two contributions $C_K$ are more explicitly written in terms of the matrix
elements over atomic $j$-$j$-coupled states\footnote{This notation is approximate since
in an external electric field $J$ is no longer strictly a good quantum number, due to the breaking of full 
rotational symmetry. However, the chosen electric field is very small, hence the
symmetry breaking is not so drastic as to lead to a breakdown of the notation. $M_J$, on the other
hand, is an exact quantum number for our zeroth-order states $\left| \psi^{(0)}_K \right>$ which are 
not perturbed by the hyperfine interaction.}
$\left| {\text{hole spinor}} \rightarrow {\text{particle spinor}}\; J,M_J \right>$
\begin{equation}
 C_1 = \frac{ \left< 5p\rightarrow 6s\; 0,0 \right| \hat{H}_{\text{HF}}
 \left| 0,0 \right> \left< 0,0 \right| 
         \imath\, \sum\limits_e\,\gamma^0_e\, \gamma^5_e\, \rho({\bf{r}}_e)
	 \left| 5p\rightarrow 6s\; 0,0 \right> } 
	 { \varepsilon_{0,0}^{(0)} - \varepsilon_{5p\rightarrow 6s\; 0,0}^{(0)} },
 \label{EQ:MAJ_ME1}
\end{equation}
\begin{equation}
 C_2 = \frac{ \left< 5p\rightarrow 6p\; 1,0 \right| \hat{H}_{\text{HF}}
 \left| 0,0 \right> \left< 0,0 \right| 
         \imath\, \sum\limits_e\,\gamma^0_e\, \gamma^5_e\, \rho({\bf{r}}_e)
	 \left| 5p\rightarrow 6p\; 1,0 \right> } 
	 { \varepsilon_{0,0}^{(0)} - \varepsilon_{5p\rightarrow 6p\; 1,0}^{(0)} },
 \label{EQ:MAJ_ME2}
\end{equation}
where $J$ and $M_J$ are $n$-body total angular momentum quantum numbers
and the Xe electronic ground state is denoted as $\left|0,0\right>$.
The off-diagonal hyperfine matrix element is larger in the term $C_2$ than in the term $C_1$. This is
reasonable, because the hyperfine Hamiltonian does not break parity symmetry (the matrix element in
$C_1$ is non-zero, however, due to $6s - 6p_{1/2}$ mixing caused by the external electric field).
Conversely, the off-diagonal S-PS-ne matrix element is larger in the term $C_1$ than in the term $C_2$.
Again, this is expected, since the S-PS-ne Hamiltonian is parity odd. Note also that the couplings
only occur between states where $\Delta M_J = M_J - M_J' = 0$, since both the hyperfine and the 
S-PS-ne Hamiltonians are rotationally invariant. So the above analysis provides
a detailed understanding of the mechanism that leads to a non-zero S-PS-ne interaction constant for
ground-state Xe in the presence of both an external electric field and magnetic hyperfine
interaction.

In analogy to similar matrix elements in the Hg atom studied in Ref. \cite{FleigJung_JHEP2018} we have
${\text{sgn}}(C_1) = -{\text{sgn}}(C_2)$. In Xe, however, ${\cal{O}}(|C_1|-|C_2|)$ is only one 
unit larger
than the order of a large number of additional small contributions (with alternating signs) which 
makes Xe a difficult case. The strategy based on this finding for obtaining a reliable value of the 
S-PS-ne interaction is explained in the following.

We determine a base value for an (uncorrelated) Random-Phase-Approximation (RPA) model to which
a correlation shift is added, determined from the comparison of a root-restricted RPA model and
the corresponding correlated model. This base value is shown in Table \ref{TAB:129Xe:AEHYEN}
for the model S8 using the cvQZ basis set. We have verified that the electron EDM enhancement $R$ with this model corresponds
well\footnote{The deviation is on the order of $10$\%. Electron EDM enhancements for closed-shell
atoms will be published separately \cite{TF2020}.} to the value obtained by 
M{\aa}rtensson-Pendrill \cite{mar-pendrill-oester_PS1987}, since S-PS-ne and electron EDM matrix elements behave
very similarly.
In order to corroborate this statement we display in Table \ref{TAB:RATIOS} the ratios of leading off-diagonal matrix elements in Xe (this work) and Hg (taken from the work for Ref. \cite{FleigJung_JHEP2018}).
                             \begin{table}[h]
                             \caption{Ratios of electron EDM \cite{TF2020} and S-PS-Ne matrix elements for leading transitions in Xe and Hg}
                             \label{TAB:RATIOS}
                                          \begin{tabular}{l|c}
                                                  Xe    &  $\left[ \frac{d_e}{C_S 10^{-18} e \text{cm}} \right]$        \\ \hline
                                                  Ratio $\frac{\left< 0,0 \right| 2\imath c\, \sum_j\, \gamma^0_j\, \gamma^5_j\, {\bf{p}}_j^2 \left| 5p\rightarrow 6s\; 0,0 \right>}{\left< 0,0 \right| \frac{AG_F}{\sqrt{2}}\imath \sum_e\,\gamma^0_e\, \gamma^5_e\, \rho({\bf{r}}_e)
                                                  \left| 5p\rightarrow 6s\; 0,0 \right>}$  &  $-158.0$    \rule{0.0cm}{0.7cm}          \\
                                                  Ratio $\frac{\left< 0,0 \right| 2\imath c\, \sum_j\, \gamma^0_j\, \gamma^5_j\, {\bf{p}}_j^2 \left| 5p\rightarrow 6p\; 1,0 \right>}{\left< 0,0 \right| \frac{AG_F}{\sqrt{2}}\imath \sum_e\,\gamma^0_e\, \gamma^5_e\, \rho({\bf{r}}_e)
                                                  \left| 5p\rightarrow 6p\; 1,0 \right>}$  &  $-159.2$    \rule{0.0cm}{0.7cm}  \\ \hline\hline
                                                  Hg    &  $\left[ \frac{d_e}{C_S 10^{-18} e \text{cm}} \right]$        \\ \hline
                                                  Ratio $\frac{\left< {^1S_{0,M_J=0}} \right| 2\imath c\, \sum_j\, \gamma^0_j\, \gamma^5_j\, {\bf{p}}_j^2 \left| {^3P_{0,M_J=0}} \right>}{\left< {^1S_{0,M_J=0}} \right| \frac{AG_F}{\sqrt{2}}\imath \sum_e\,\gamma^0_e\, \gamma^5_e\, \rho({\bf{r}}_e)
                                                  \left| {^3P_{0,M_J=0}} \right>}$  &  $-85.9$    \rule{0.0cm}{0.7cm}          \\
                                                  Ratio $\frac{\left< {^1S_{0,M_J=0}} \right| 2\imath c\, \sum_j\, \gamma^0_j\, \gamma^5_j\, {\bf{p}}_j^2 \left| {^3S_{1,M_J=0}} \right>}{\left< {^1S_{0,M_J=0}} \right| \frac{AG_F}{\sqrt{2}}\imath \sum_e\,\gamma^0_e\, \gamma^5_e\, \rho({\bf{r}}_e)
                                                  \left| {^3S_{1,M_J=0}} \right>}$  &  $-85.2$    \rule{0.0cm}{0.7cm}       
                                          \end{tabular}
                                  \end{table}
These ratios are very stable across different transitions in a respective atom with a given nucleon number. Moreover, their absolute values line up extremely well with the ratios for matrix elements in paramagnetic atoms given by Dzuba {\etal} in Ref. \cite{PhysRevA.84.052108,PhysRevA.85.029901}.

\begin{table}[h]

\caption{\label{TAB:129Xe:AEHYEN}
    S-PS-ne interaction ratio $S$ and interaction constant $\alpha_{C_S}$ in ${^1S}_0$ ground state 
    of the {$^{129}$Xe} isotope, $I=1/2$, $g({^{129}{\text{Xe}}}) = -0.777977$, $E_{\text{Ext}} = 0.0005$ \au,
    for sets of perturbation CI states 
    (all CI model X-SDT8\_SD8, except for $^*$CI model S8-X-SDT8\_SD16, $^{**}$CI model S10-X-SDT8\_SD18,
    and $^Q$ CI model X-SDTQ8\_SD8)
         }

 \vspace*{-0.4cm}
 \hspace*{-1.7cm}
 \begin{center}
 \begin{tabular}{l|l|cc}
 Basis & \# of CI states/X  & $S$ [$10^{-3}$ \au] & $\alpha_{C_S}$ [$10^{-23}\, e$ cm] \\ \hline
  cvTZ/40 a.u. & $8$/S8                           &  $ 0.590$                & $ 0.633$     \\
  cvTZ/7 a.u.  & $8$/6s6p                         &  $ 0.552$                & $ 0.592$     \\
  cvTZ/14 a.u. & $8$/6s6p                         &  $ 0.554$                & $ 0.594$     \\
  cvTZ/7 a.u.  & $8$/6s6p5d                       &  $ 0.604$                & $ 0.648$     \\
  cvTZ/7 a.u.  & $8$/6s6p5d$^Q$                   &  $ 0.603$                & $ 0.647$     \\
  cvTZ/7 a.u.  & $8$/6s6p5d7p                     &  $ 0.667$                & $ 0.716$     \\
  cvTZ/7 a.u.  & $8$/6s6p5d7p7s                   &  $ 0.693$                & $ 0.744$     \\
  cvTZ/14 a.u. & $8$/6s6p5d7p7s                   &  $ 0.694$                & $ 0.745$     \\
  cvTZ/14 a.u. & $8$/6s6p5d7p7s$^*$               &  $ 0.731$                & $ 0.784$     \\
  cvTZ/14 a.u. & $8$/6s6p5d7p7s$^{**}$            &  $ 0.717$                & $ 0.769$     \\
  cvTZ/7 a.u.  & $8$/6s6p5d7p7s6d                 &  $ 0.699$                & $ 0.750$     \\
  cvTZ/7 a.u.  & $8$/6s6p5d7p7s6d4f               &  $ 0.702$                & $ 0.753$     \\
  cvTZ/7 a.u.  & $8$/6s6p5d7p7s6d4f7d5f8p8s       &  $ 0.671$                & $ 0.720$     \\ 
  cvTZ/7 a.u.  & $8$/all                          &  $ 0.583$                & $ 0.625$     \\ \hline
  cvQZ/100 a.u. & $8$/S8                          &  $ 0.592$                & $ 0.635$     \\
  cvQZ/50 a.u.  & $1000$/S8                       &  $ 0.499$                & $ 0.535$     \\
  cvQZ/100 a.u. & $1281$/S8                       &  $ 0.611$                & $ 0.655$     \\
  cvQZ/50 a.u.  & $8$/6s6p                        &  $ 0.482$                & $ 0.517$     \\
  cvQZ/50 a.u.  & $8$/6s6p5d7p7s                  &  $ 0.710$                & $ 0.762$     \\ \hline
 {\bf{vQZ/1281/S8/100}} + $\boldsymbol{\Delta S_{\text{corr}}}$ & & $\boldsymbol{0.680}$ & $\boldsymbol{0.730}$
 \end{tabular}
 \end{center}
\end{table}

We first notice that the basis set effect onto the two main contributors is negligibly small
(compare the two $8$/S8 models). It is an intriguing observation that the correlation correction
from the minimal correlation model in TZ basis ($8$/6s6p) is negative (\ie{} it makes the $S$
ratio smaller than the reference RPA value), then becomes positive when the spinor set
into which triple excitations are allowed is increased, and finally converges to a value that
is very close to the RPA value when the full set of spinors below $7$\au{} is opened for triple
excitations.

Based on these findings we determine the final correlation correction from corresponding
calculations, determined as follows:
\begin{eqnarray*}
\Delta S_{\text{corr}} &:=& S({\rm{vTZ/8/all/7}}) - S({\rm{vTZ/8/S8/40}}) \\
		       &+&  S({\rm{vQZ/8/6s6p5d7p7s/50}}) - S({\rm{vTZ/8/6s6p5d7p7s/14}}) \\
		       &+&  S({\rm{vTZ/8/6s6p5d7p7s/14^*}}) - S({\rm{vTZ/8/6s6p5d7p7s/14}}) \\
		       &+&  S({\rm{vTZ/8/6s6p5d7p7s/14^{**}}}) - S({\rm{vTZ/8/6s6p5d7p7s/14}}) \\
				     &\approx& 0.069 \times 10^{-3} \au
\end{eqnarray*}
Note that for models of type S8 the virtual truncation is arbitrary since here there is no correlating
spinor space as in all other models. The final value for the S-PS-ne interaction ratio is obtained by
adding the correlation correction to the best present RPA result, as shown in Table \ref{TAB:129Xe:AEHYEN}.
The models denoted with asterisks correct for core-valence correlations among the $4s4p$ and the
valence electrons ($^*$) and among the $4d$ and the valence electrons ($^{**}$).
These latter corrections comply with what is expected in physical terms: Valence correlations give
the largest correction, followed by correlations with outer-core $s$ and $p$ electrons, followed 
by correlations with outer-core $d$ electrons.

The uncertainties of the final value for the atomic S-PS-ne interaction constant are $2$\% for the 
atomic basis set, $5$\% for remaining inner-shell correlations, $1$\% for the higher excitation ranks
not considered by the present models and $16$\% for the lack of correlation corrections to other
contributing states, adding up (linearly) to a total estimated uncertainty of about $25$\%.
Thus, our interaction constant is given as
\begin{equation}
 \alpha_{C_S} = \left( 0.71 \pm 0.18 \right) [10^{-23}\, e\, \text{cm}].
 \label{eq::CSres}
\end{equation}
The principal physical effects neglected in the above calculations are electron correlation
effects on the ensemble of individually minor contributions to $\alpha_{C_S}$. These minor contributions are accounted
for, however, at the RPA level of theory.
\section{Phenomenology}
\label{SEC:PHENO}
Recently there has been significant progress in the experimental techniques for determining the EDM of ${}^{129}$Xe
\cite{Allmendinger:2019jrk,Sachdeva:2019rkt,Sachdeva:2019blc,2018PhLA..382..588S,Liu:2020lhw}, with the strongest upper limit
now reading \cite{Liu:2020lhw}
\begin{equation}\label{eq::dXeexp}
|d_{\rm Xe}|\leq 7.4\times10^{-28}e\,{\rm cm}\quad (95\%~{\rm CL,~preliminary})\,.
\end{equation}
While the resulting limits on
phenomenological parameters are still relatively weak, there is the potential to improve the sensitivity by several
orders of magnitude in the mid-term future. Even with these improvements, Xenon will likely not provide the strongest
limits on ${\cal{CP}}$-violating parameters in single-source analyses, 
given its
lower sensitivity to these parameters by approximately a factor of 10 compared to ${}^{199}$Hg.
However, limits obtained from single-source analyses cannot be considered reliable, since they require neglecting potential 
cancellations. Given that in virtually every BSM model several ${\cal{CP}}$-violating sources at low energy exist, such an
assumption can usually not be justified and results in a mere estimate of the order of magnitude for a given limit.
Generalizing the analysis to include several sources of ${\cal{CP}}$ violation comes, however, at the cost that a given
measurement does not constrain any individual parameter at all, only a system-specific linear combination. The
phenomenological impact especially of additional diamagnetic systems like ${}^{129}$Xe is then to provide independent
linear combinations in such a general approach, which is necessary to eventually provide limits on individual
coefficients and achieve some level of model-discrimination.
Diamagnetic systems can also play a role in determining the
electron EDM in this broader context \cite{Jung:2013mg,FleigJung_JHEP2018}. Finally, from a more theoretical point of view
the calculated quantities are of interest in order to test approximate relations that are commonly used to estimate
specifically $\alpha_{C_S}$ from a calculation of $\alpha_{C_T}$
\cite{Flambaum_Khriplovich1985,Kozlov:1988qn,khriplovich_lamoreaux} for a variety of systems. This provides unique insight into the precision of these
approximations, since this is the first explicit calculation of both quantities within the same framework. In the
following we discuss these points in more detail.

\subsection{Xenon-EDM in generalized fits}
In Refs.~\cite{Jung:2013mg,FleigJung_JHEP2018} it was shown how EDMs of diamagnetic systems can be used to model-independently
constrain the electron EDM and the S-PS nucleon-electron interaction beyond what is presently possible with
paramagnetic systems alone, since all of the latter constrain similar linear combinations of these two quantities. In
principle, Xenon could be helpful: our calculation shows that the coefficients of $d_e$ and $C_S$ exhibit
opposite signs (assuming the sign of the calculation in Ref.~\cite{mar-pendrill-oester_PS1987} is correct), in contrast
to all paramagnetic systems, but similarly to Mercury. However, given that the limit in Eq.~\eqref{eq::dXeexp} is still
much weaker than the one on the Mercury EDM \cite{Heckel_Hg_PRL2016}, and an overall reduced sensitivity, this option is
rendered less interesting; in fact, repeating the analysis from Ref.~\cite{FleigJung_JHEP2018} including the new result
for ThO from the ACME collaboration \cite{ACME_ThO_eEDM_nature2018}, we conclude that the corresponding contribution to
Xenon is limited as
\begin{equation}\label{eq::dXeconstr}
|d_{\rm Xe}(d_e,C_S)|\leq 10^{-30}e\,{\rm cm}\quad\mbox{($95\%$ CL, from paramagnetic systems plus Hg)}\,.
\end{equation}
This result uses our calculation for $\alpha_{C_S}$, together with the estimate for $\alpha_{d_e}$ from
\cite{mar-pendrill-oester_PS1987}, with $100\%$ uncertainty assigned to the latter.

This observation can be turned into a virtue: we consider a scenario in which the main contributions to low-energy EDMs
stem from semileptonic operators (the electron EDM can be added without changing the argument). Such a scenario is
broadly motivated by present indications for lepton-non-universal BSM physics in both charged-current ($b\to c\tau\nu$) and
neutral-current ($b\to s\ell\ell$) decays \cite{Amhis:2019ckw}. While a connection to EDMs is not implied at all by
these anomalies, specific explanations are constrained by EDMs, in particular some leptoquark models
\cite{Becirevic:2018afm,Dekens:2018bci,Altmannshofer:2020ywf}. Given the constraint in Eq.~\eqref{eq::dXeconstr},
together with the observation that a pseudoscalar-scalar contribution to diamagnetic EDMs might be sizable, but not
larger than the contribution from $C_S$ \cite{Ellis:2008zy,Jung:2013hka}, the limit on $d_{\rm Xe}$ would imply in such
a scenario a limit on the combination of $C_T^p,C_T^n$ relevant for Xenon. 
This combination is not precisely known, however; while in semi-empirical models \cite{1937ZPhy..106..358S,Engel:1989ix,Flambaum:2006ip} the ratio $\langle \sigma_p\rangle/\langle\sigma_n\rangle$ can reach values larger than $1/3$, it has been argued in Ref.~\cite{Kimball_2015} that the assumptions on which these estimates are based do not hold for Xenon. 
Large-scale shell model calculations \cite{Ressell:1997kx,Toivanen:2009zza,Klos:2013rwa,Yoshinaga:2014eaa} indicate much smaller values, still with large uncertainties. For Mercury both semi-empirical estimates \emph{and} simulations \cite{Yanase:2020agg} indicate a small value of $\langle \sigma_p\rangle/\langle\sigma_n\rangle\lesssim 10\%$.
Should these coefficients turn out to be sufficiently different, the two together could
determine both contributions to $C_T$ and thereby $C_T$ for any other system. Using the above argument for Mercury to
isolate the $C_T$ contribution would require sufficient control on $C_S$ from paramagnetic systems alone, which should be attainable in the future. 

Explicitly, we obtain in this scenario 
from combining the experimental result in Eq.~\eqref{eq::dXeexp} with our calculation for $\alpha_{C_T}$ (Eq.~\eqref{eq::CTres}) and 
the estimate \cite{Ressell:1997kx,Toivanen:2009zza,Klos:2013rwa}
$\langle\Sigma\rangle_{\rm Xe} \sim 0.6$ 
\begin{equation}
|C_T^{\rm Xe}|\leq 2.6\times 10^{-7}\quad (95\%~{\rm CL)\,.}
\end{equation}

\subsection{Testing semi-analytical relations}
The S-PS nucleon-electron interaction constant is commonly estimated using a phenomenological relation to the
coefficient of the \PT-odd tensor interaction $\alpha_{C_T}$
\cite{Flambaum_Khriplovich1985,Kozlov:1988qn,khriplovich_lamoreaux}. Using
\begin{equation}
\alpha_{C_T} = 10^{-20}C_{C_T}\frac{\langle\boldsymbol{\Sigma}\rangle\cdot \mathbf I}{I}e\,{\rm cm}\,,
\end{equation}
this can be cast into the following form:
\begin{align}\label{eq::CSCT}
\alpha_{C_S} = 5.3\times 10^{-24}(1+0.3\,Z^2\alpha^2)A^{2/3}\mu_A C_{C_T} e\,{\rm cm}\equiv r_{S/T} C_{C_T}\,,
\end{align}
where $\mu_A$ denotes the magnetic moment of the atom's nucleus (in units of the nuclear magneton).
This relation implies a ratio $r_{S/T}^{\rm Xe} = -6.6
\times 10^{-23}e\,{\rm cm}$, to be compared to
$r_{S/T}^{\rm Xe}= (1.29\pm0.35)\times 10^{-23}e\,{\rm cm}$ 
obtained in our explicit calculation. Neither
modulus nor sign are hence reproduced by the relation \eqref{eq::CSCT}. Performing the same test with Mercury, using our
recent calculations \cite{FleigJung_JHEP2018,Fleig_PRA2019}, yields $r_{S/T}^{\rm Hg}=1.01\times 10^{-22}e\,{\rm cm}$
from relation \eqref{eq::CSCT}, to be contrasted with $r_{S/T}^{\rm Hg} = (0.63\pm 0.15)\times 10^{-22}e\,{\rm cm}$ from
explicit calculation. In this case the sign agrees and also the modulus is better reproduced, although it is still $\sim 40\%$ apart. This leads us to
conclude that Eq.~\eqref{eq::CSCT} should not be used for anything else than a rough order-of-magnitude estimate, while
for quantitative results a dedicated calculation is necessary. Note that our present calculations are consistent with those of Ref.~\cite{mar-pendrill-oester_PS1987} concerning the signs of the relevant atomic interaction constants. Furthermore, we improve upon the methods used there by including leading electron correlation corrections.

There is another relation between the coefficients for $C_T$ and the electron dipole moment
\cite{Flambaum_Khriplovich1985,Kozlov:1988qn,khriplovich_lamoreaux}. It shows a similar failure as the relation above.
Since the matrix elements entering the coefficients of $C_S$ and $d_e$ behave similarly, one could consider
eliminating $C_T$ from these equations to obtain a more reliable relation. However, even in this case the sign issue
remains: the coefficients of $d_e$ and $C_S$ are predicted to have strictly the same sign, while the opposite is true
in explicit calculations.
%
%
\section{Conclusions}
\label{SEC:CONCL}
We calculate the interaction constants for the \PT-odd operators with tensor-pseudotensor and scalar-pseudoscalar structures, obtaining
\begin{equation}
 \alpha_{C_T} = \left( 0.520 \pm 0.049 \right) [10^{-20}\, \left<\Sigma\right>_{\text{Xe}}\, e\, \text{cm}]\quad\mbox{and}\quad \alpha_{C_S} = \left( 0.71 \pm 0.18 \right) [10^{-23}\, e\, \text{cm}].
 \end{equation}
 The former value is in good agreement with existing calculations, see table~\ref{TAB:RT_XE}, while the latter is the first explicit calculation of this quantity. These calculations help relating fundamental ${\cal{CP}}$-violating parameters to present and future measurements of the EDM of ${}^{129}$Xe, which will help to disentangle different BSM effects contributing to EDMs.
 
 We applied our calculations in the extraction of a limit for the BSM coefficient $C_T^{\text{Xe}}$ from present data, obtaining $|C_T^{\text{Xe}}|\leq 2.6\times 10^{-7}$ at 95\% CL, holding not only in a single-source analysis, but also when allowing for the presence of \PT-odd semileptonic operators and the electron EDM. While this limit seems much weaker than the one obtained for the corresponding quantity in Mercury, it is presently not possible to model-independently infer a stronger bound from the latter, since in general a different linear combination of $C_T^p$ and $C_T^n$ enters. We furthermore used our calculation to test explicitly relation~\eqref{eq::CSCT}: while the relation is reasonably well fulfilled for Mercury, we find that it reproduces neither the correct sign nor order of magnitude of our result. We conclude that for quantitative analyses this phenomenological relation is insufficient and  explicit calculations are necessary. We are looking forward to improved measurements of the Xenon EDM in the coming years which will become an important ingredient in global analyses of EDMs.
 
\acknowledgments
\label{SEC:ACK}
We thank Micka\"el Hubert (Toulouse) for making Gaussian-expanded Fermi densities for nuclei available prior to publication. We also thank Vladimir Dzuba (Sydney) and Victor Flambaum (Sydney) for helpful discussions.
This research was supported by the DFG cluster of excellence “Origin and Structure of the Universe”.
The work of MJ is supported by the Italian Ministry of Research (MIUR) under grant PRIN 20172LNEEZ.
\begin{appendix}

\section{Closed-shell State Expectation Values}
\label{APP:B}

We consider an electronically closed-shell Slater determinant for an atomic configuration $ns_{1/2}^2$ and a corresponding determinant
for a configuration $np_{1/2}^2$. The four one-particle spinors constituting these determinants can be written \cite{bethe} as
\begin{equation}
	u_{s(m_j=1/2)} = \left( \begin{array}{c} g_s(r)\, Y_{0,0} \\ 0 \\ -\frac{\imath}{\sqrt{3}} f_s(r) Y_{1,0} \\ -\imath\sqrt{\frac{2}{3}} f_s(r) Y_{1,1}
	\end{array} \right) \,, \hspace{1.0cm}
	u_{s(m_j=-1/2)} = \left( \begin{array}{c} 0 \\ -g_s(r)\, Y_{0,0} \\ -\imath\sqrt{\frac{2}{3}} f_s(r) Y_{1,-1} \\ -\frac{\imath}{\sqrt{3}} f_s(r) Y_{1,0} 
		              \end{array} \right)
\label{AEQ:SSPI}
\end{equation}
for orbital angular momentum $\ell=0$ and
\begin{equation}
	u_{p(m_j=1/2)} = \left( \begin{array}{c} \frac{1}{\sqrt{3}} g_p(r)\, Y_{1,0} \\ \frac{2}{3} g_p(r)\, Y_{1,1} \\ -\imath f_p(r) Y_{0,0} \\ 0
	\end{array} \right) \,, \hspace{1.0cm}
	u_{p(m_j=-1/2)} = \left( \begin{array}{c} \frac{2}{3} g_p(r)\, Y_{1,-1} \\ \frac{1}{\sqrt{3}} g_p(r)\, Y_{1,0} \\ 0 \\ \imath f_p(r) Y_{0,0} 
		              \end{array} \right)
\label{AEQ:PSPI}
\end{equation}
for orbital angular momentum $\ell=1$, where $Y_{\ell,m_{\ell}}$ are the spherical harmonics and $f,g$ are radial functions.

In the presence of a constant and uniform external electric field in $z$ direction a perturbative Hamiltonian is written as
\begin{equation}
	{\hat{H}}_z = \hat{z} E_{z,{\text{ext}}}\,,
\end{equation}
and the multiplicative operator is expressed in spherical polar coordinates as $z = r \cos\vartheta$. For reasons of rotational symmetry
the only spinor from the set in Eqs. (\ref{AEQ:SSPI}) and (\ref{AEQ:PSPI}) that will be mixed into the ground-state spinor $u_{s(m_j=1/2)}$
by ${\hat{H}}_z$ is $u_{p(m_j=1/2)}$. The mixing coefficient is a function of the coupling matrix element
\begin{eqnarray}
	\nonumber
	\left< u_{s(1/2)} \left| r \cos\vartheta \right| u_{p(1/2)} \right> &=&
	   \frac{1}{\sqrt{3}} \left< g_s(r) |r^3| g_p(r) \right> \left< Y_{0,0} |\sin\vartheta \cos\vartheta| Y_{1,0} \right> + 0 \\
	\nonumber
	   && 
	  +\frac{1}{\sqrt{3}} \left< f_s(r) |r^3| f_p(r) \right> \left< Y_{1,0} |\sin\vartheta \cos\vartheta| Y_{0,0} \right> + 0 \\
	  &=& \frac{1}{3} \left[ \left< g_s(r) |r^3| g_p(r) \right> + \left< f_s(r) |r^3| f_p(r) \right> \right]\,,
\end{eqnarray}
where the two integrations $<,>$ are purely radial/spherical, respectively, and the spherical volume element 
$d{\cal{V}} = r^2 \sin\vartheta dr d\vartheta d\varphi$ has been assumed. The mixing coefficient can thus be written as
\begin{equation}
	c_{1/2}(E) = \frac{\frac{1}{3} \left[ \left< g_s(r) |r^3| g_p(r) \right> + \left< f_s(r) |r^3| f_p(r) \right> \right]\, E_{z,{\text{ext}}}}
	            {\Delta \varepsilon}\,,
\end{equation}
where $\Delta \varepsilon$ is the energy splitting between the $s_{1/2}$ and $p_{1/2}$ levels. Likewise, there is only one mixing matrix
element for the other ground-state spinor $u_{s(m_j=-1/2)}$
\begin{eqnarray}
        \nonumber
        \left< u_{s(-1/2)} \left| r \cos\vartheta \right| u_{p(-1/2)} \right> &=&
	   -\frac{1}{3} \left[ \left< g_s(r) |r^3| g_p(r) \right> + \left< f_s(r) |r^3| f_p(r) \right> \right]\,,
\end{eqnarray}
and the corresponding mixing coefficient is
\begin{equation}
	c_{-1/2}(E) = -\frac{\frac{1}{3} \left[ \left< g_s(r) |r^3| g_p(r) \right> + \left< f_s(r) |r^3| f_p(r) \right> \right]\, E_{z,{\text{ext}}}}
	            {\Delta \varepsilon} = -c_{1/2}(E)\,.
\end{equation}
Based on these results we can write $E$-field-perturbed\,, unnormalized ground-state spinors as
\begin{eqnarray}
	u_+ &=& u_{s(1/2)} + c_{1/2}(E)\, u_{p(1/2)}\,, \\
	u_- &=& u_{s(-1/2)} - c_{1/2}(E)\, u_{p(-1/2)}.
\end{eqnarray}
Closed-shell determinants from spinors with even orbital angular momentum quantum number $\ell$ cannot contribute due to parity symmetry.
Those from  spinors with odd orbital angular momentum quantum number $\ell \ge 3$ cannot contribute, either, because the electric
dipole operator $\hat{\bf{r}}$ is only a rank-1 tensor operator.

In this two-state two-body model the antisymmetrized closed-shell wavefunction is then
\begin{equation}
	\psi_{1,2}(E) = \frac{1}{\sqrt{2}}\, \left[ u_+(1)u_-(2) - u_-(1)u_+(2) \right]
\end{equation}
for particles ``1'' and ``2''.

\subsection{${\boldsymbol{\Sigma}}$ Expectation Values}
\label{ASEC:SIGMA}

The electronic Dirac spin matrix for two electrons is ${\boldsymbol{\Sigma}} = {\boldsymbol{\Sigma}}(1) + {\boldsymbol{\Sigma}}(2)$. 
The $z$ component of the spin expectation value in the state $\psi_{1,2}(E)$ is then written out as
\begin{eqnarray}
	\nonumber
	\hspace*{-1.0cm}
	\big< \psi_{1,2}(E) \left| \left( \Sigma_z(1) + \Sigma_z(2) \right) \right| \psi_{1,2}(E) \big> &=&
	   \frac{1}{2}\, \bigg[ \left< u_+(1) \left| \Sigma_z(1) \right| u_+(1) \right> \left< u_-(2) | u_-(2) \right>  \\
	\nonumber
	   &&                 - \left< u_+(1) \left| \Sigma_z(1) \right| u_-(1) \right> \left< u_-(2) | u_+(2) \right>  \\
	\nonumber
	   &&                 - \left< u_-(1) \left| \Sigma_z(1) \right| u_+(1) \right> \left< u_+(2) | u_-(2) \right>  \\
	\nonumber
	   &&                 + \left< u_-(1) \left| \Sigma_z(1) \right| u_-(1) \right> \left< u_+(2) | u_+(2) \right>  \\
	\nonumber
 	&&                    + \left< u_+(2) \left| \Sigma_z(2) \right| u_+(2) \right> \left< u_-(1) | u_-(1) \right>  \\
	\nonumber
	   &&                 - \left< u_+(2) \left| \Sigma_z(2) \right| u_-(2) \right> \left< u_-(1) | u_+(1) \right>  \\
	\nonumber
	   &&                 - \left< u_-(2) \left| \Sigma_z(2) \right| u_+(2) \right> \left< u_+(1) | u_-(1) \right>  \\
	   &&                 + \left< u_-(2) \left| \Sigma_z(2) \right| u_-(2) \right> \left< u_+(1) | u_+(1) \right>  \bigg]\,.
	   \label{AEQ:PSI12}
\end{eqnarray}
For the overlap integrals we calculate straightforwardly
\begin{eqnarray}
	\nonumber
	\left< u_+ | u_+ \right> &=& \left< u_{s(1/2)} | u_{s(1/2)} \right> + c_{1/2}(E) \left< u_{s(1/2)} | u_{p(1/2)} \right> \\
	\nonumber
	&&      + c^*_{1/2}(E) \left< u_{p(1/2)} | u_{s(1/2)} \right> + |c_{1/2}(E)|^2 \left< u_{p(1/2)} | u_{p(1/2)} \right>  \\
	\nonumber
	&=&  \left< g_s(r) |r^2| g_s(r) \right> + \left< f_s(r) |r^2| f_s(r) \right> \\
	&&     + |c_{1/2}(E)|^2 \big( \left< g_p(r) |r^2| g_p(r) \right> + \left< f_p(r) |r^2| f_p(r) \right> \big) \\
	\label{AEQ:OVERLAP1}
	&=& \left< u_- | u_- \right> \\
	\left< u_+ | u_- \right> &=& \left< u_- | u_+ \right> = 0\,.
	\label{AEQ:OVERLAP2}
\end{eqnarray}
For the spin integrals we obtain
\begin{eqnarray}
	\nonumber
	\left< u_+ |\Sigma_z| u_+ \right> &=& \left< u_{s(1/2)} |\Sigma_z| u_{s(1/2)} \right> + c_{1/2}(E) \left< u_{s(1/2)} |\Sigma_z| u_{p(1/2)} \right> \\
	\nonumber
	&&      + c^*_{1/2}(E) \left< u_{p(1/2)} |\Sigma_z| u_{s(1/2)} \right> + |c_{1/2}(E)|^2 \left< u_{p(1/2)} |\Sigma_z| u_{p(1/2)} \right>  \\
	\nonumber
	&=&  \left< g_s(r) |r^2| g_s(r) \right> -\frac{1}{3} \left< f_s(r) |r^2| f_s(r) \right> \\
	\label{AEQ:SPININT1}
	&&     + |c_{1/2}(E)|^2 \big( -\frac{1}{3}\left< g_p(r) |r^2| g_p(r) \right> + \left< f_p(r) |r^2| f_p(r) \right> \big)\,, \\
	\nonumber
	\left< u_- |\Sigma_z| u_- \right> 
	&=&  -\left< g_s(r) |r^2| g_s(r) \right> +\frac{1}{3} \left< f_s(r) |r^2| f_s(r) \right> \\
	&&     + |c_{1/2}(E)|^2 \big( \frac{1}{3}\left< g_p(r) |r^2| g_p(r) \right> - \left< f_p(r) |r^2| f_p(r) \right> \big)\,.
	\label{AEQ:SPININT2}
\end{eqnarray}
Using the results from Eqs. (\ref{AEQ:OVERLAP1}), (\ref{AEQ:OVERLAP2}), (\ref{AEQ:SPININT1}) and (\ref{AEQ:SPININT2}), Eq.~(\ref{AEQ:PSI12}) becomes
\begin{eqnarray}
	\nonumber
	\big< \psi_{1,2}(E) \left| \left( \Sigma_z(1) + \Sigma_z(2) \right) \right| \psi_{1,2}(E) \big> &=& \\
	&& \hspace*{-4.0cm}  2 \frac{1}{2}\, \big( \left< u_+ \left| \Sigma_z \right| u_+ \right>  + \left< u_- \left| \Sigma_z \right| u_- \right> \big) \left< u_+ | u_+ \right> = 0\,.
\end{eqnarray}
Since in the matrices $\Sigma_x$ and $\Sigma_y$ the Pauli matrices $\sigma_x$ and $\sigma_y$ are purely off-diagonal, it is easily
demonstrated that 
\begin{eqnarray*}
	\big< \psi_{1,2}(E) \left| \left( \Sigma_x(1) + \Sigma_x(2) \right) \right| \psi_{1,2}(E) \big> &=& 0\,, \\
	\big< \psi_{1,2}(E) \left| \left( \Sigma_y(1) + \Sigma_y(2) \right) \right| \psi_{1,2}(E) \big> &=& 0\,,
\end{eqnarray*}
essentially due to vanishing large- and small-component overlaps $\left< u_{L_1} | u_{L_2} \right> = \left< u_{S_1} | u_{S_2} \right> = 0$,
irrespective of the spinor type $u$.

\subsection{$\boldsymbol{\gamma_3}$ Expectation Values}
\label{ASEC:GAMMA}

Up to Eq. (\ref{AEQ:OVERLAP2}) the expressions remain identical. Then we calculate
\begin{eqnarray}
	\nonumber
	\left< u_+ |\gamma_3| u_+ \right> &=& c_{1/2}(E) \big[ -\imath \left< g_s(r) |r^2| f_p(r) \right> -\frac{\imath}{3} 
	      \left< f_s(r) |r^2| g_p(r) \right> + \imath \frac{2}{3} \left< f_s(r) |r^2| g_p(r) \right> \big] \\
	\nonumber
	 && + c^*_{1/2}(E) \big[ -\frac{\imath}{3} \left< g_p(r) |r^2| f_s(r) \right> + \imath \frac{2}{3} \left< g_p(r) |r^2| f_s(r) \right>
	    - \imath \left< f_p(r) |r^2| g_s(r) \right> \big] \\
	 &=& c_{1/2}(E) \big[ \imath \frac{2}{3} \left< g_p(r) |r^2| f_s(r) \right> - 2\imath \left< f_p(r) |r^2| g_s(r) \right> \big]\,, \\
	\nonumber
	\left< u_- |\gamma_3| u_- \right> &=& 
	    c_{1/2}(E) \big[ \imath \frac{2}{3} \left< g_p(r) |r^2| f_s(r) \right> - 2\imath \left< f_p(r) |r^2| g_s(r) \right> \big]
	    = \left< u_+ |\gamma_3| u_+ \right>\,,
\end{eqnarray}
and so the total expectation value becomes
\begin{eqnarray}
	\nonumber
	\big< \psi_{1,2}(E) \left| \left( \gamma_3(1) + \gamma_3(2) \right) \right| \psi_{1,2}(E) \big> =
	  2 \frac{1}{2}\, \big( \left< u_+ \left| \gamma_3 \right| u_+ \right>  + \left< u_- \left| \gamma_3 \right| u_- \right> \big) \left< u_+ | u_+ \right>  \\
	\nonumber
	  && \hspace*{-12.5cm} = 2 c_{1/2}(E) \big[ \imath \frac{2}{3} \left< g_p(r) |r^2| f_s(r) \right> - 2\imath \left< f_p(r) |r^2| g_s(r) \right> \big] \\
	\nonumber
	&& \hspace*{-11.5cm} \bigg[ \left< g_s(r) |r^2| g_s(r) \right> + \left< f_s(r) |r^2| f_s(r) \right>  \\
	&& \hspace*{-10.5cm}   + |c_{1/2}(E)|^2 \big( \left< g_p(r) |r^2| g_p(r) \right> + \left< f_p(r) |r^2| f_p(r) \right> \big) \bigg]\,,
\end{eqnarray}
which is non-zero.

\end{appendix}
\bibliographystyle{unsrt}
\newcommand{\Aa}[0]{Aa}

\clearpage
\end{document}